\documentclass[aps,pra,twocolumn,superscriptaddress,showpacs]{revtex4}
\usepackage{hyperref}

\usepackage{graphicx}

\usepackage{amsmath}
\usepackage{epsfig}

\begin{document}
\title{Environmental Assisted Quantum Information Correction for Continuous Variables}

\author{Metin Sabuncu}\email{metin.sabuncu@mpl.mpg.de}
\affiliation{Department of Physics,
Technical University of Denmark, 2800 Kongens Lyngby, Denmark}
\affiliation{Max Planck Institute for the Science of Light, G\"unther Scharowskystrasse 1, 91058 Erlangen, Germany}
\author{Radim Filip}
\affiliation{Department of Optics, Palack\' y
University, 17. listopadu 50,  772~07 Olomouc, Czech Republic}
\author{Gerd Leuchs}
\affiliation{Max Planck Institute for the Science of Light, G\"unther Scharowskystrasse 1, 91058 Erlangen, Germany}
\affiliation{University Erlangen-N\"urnberg, Staudtstrasse 7/B2, 91058 Erlangen, Germany}
\author{Ulrik L. Andersen}
\affiliation{Department of Physics,
Technical University of Denmark, 2800 Kongens Lyngby, Denmark}

\date{\today}

\begin{abstract}
Quantum information protocols are inevitably affected by decoherence which is associated with the leakage of quantum information into an environment. In this paper we address the possibility of recovering the quantum information from an environmental measurement. We investigate continuous variable quantum information, and we propose a simple environmental measurement that under certain circumstances fully restores the quantum information of the signal state although the state is not reconstructed with unit fidelity. We implement the protocol for which information is encoded into conjugate quadratures of coherent states of light and the noise added under the decoherence process is of Gaussian nature. The correction protocol is tested using both a deterministic as well as a probabilistic strategy. The potential use of the protocol in a continuous variable quantum key distribution scheme as a means to combat excess noise is also investigated.
\end{abstract}


\maketitle
\section{Introduction}

Quantum communication  is fundamentally more secure than traditional classical communication as exemplified by quantum key distribution \cite{bennett}. This fact alone has recently triggered a lot of research and development in quantum communication \cite{gisin}. However the fragile nature of quantum signals makes the realization of many quantum communication tasks far from trivial. The state carrying quantum information interacts inevitably with the surrounding environment, thus introducing decoherence into the system and consequently leakage of information into the environment~\cite{zurek}. This trend pushes one toward the classical regime making decoherence the main obstacle in quantum information protocols.

Decoherence of quantum information occurs in all real life quantum communication links such as fibers, free space propagation and quantum memories~\cite{rarity}.
However, there exist techniques which are capable of maintaining the "quantumness" of the signals as they propagate through noisy environments. One of them is quantum error correcting coding where the quantum information is encoded in a subspace of a large and more robust Hilbert space to prevent decoherence~\cite{shor95.pra,steane96.prl}. Another well known method is that of entanglement purification combined with teleportation~\cite{bennett2}. If the noise in the channel is Gaussian and the information carrying quantum states are Gaussian, the methods of quantum error correction coding and entanglement distillation will rely on experimentally challenging non-Gaussian operations~\cite{niset08,eisert02.prl,fiurasek02.prl,giedke02.pra}.

In all of the above mentioned schemes, one assumes that one will have no access to the information that has leaked into the environmental quantum system. If, however, one has some control over the external system (the environment) it is actually possible to reverse the devastating interaction of the environment through measurements and classical feedforward even for the case where the noise in the environment is Gaussian. This has been already demonstrated in a technique called quantum erasing: The information that has leaked into the external system is erased through a special tailored measurement and the quantum state is subsequently fully restored by apropriate feedforward~\cite{erasing}. The idea of quantum erasing was first implemented using single
photons \cite{Schwindt,Tsegaye} and later  extended to the continuous
variable regime in Ref.~\cite{Filip1}, and demonstrated using squeezed light~\cite{Andersen}. It has also been shown that by employing the technique of erasing with squeezed light, a potential loss of continuous variable quantum information can be corrected \cite{Marek}.

Quantum erasing has been discussed as a
method of environmental assisted {\em quantum state} correction for a
qubit (qutrit) channel \cite{Werner}. Remarkably, it was found that any random-unitary CP map
is invertible by quantum erasing. It means that any noise or errors in qubit and
qutrit carrying channels can be overcome by using the trick of quantum erasing, that is, by measuring the external system (the environment) and subsequently correct the quantum state based on the measurement outcomes. The correction of a qubit as a result of an environental measurement was recently demonstrated experimentally~\cite{Sciarrino}. On the other hand, it was proved that for states described in higher dimensional Hilbert spaces, such as continuous variable states, perfect state
reconstruction is generally not possible (even if the environment is in a pure state)~\cite{Buscemi1}. There are however special cases where a reversible interaction for continuous variables is possible. For example, if the modes of the environment interacting with the signal are quadrature squeezed, a measurement of the anti-squeezed quadrature of the environmental modes after interaction may enable a perfect recovery of the otherwise deteroriated continuous variable state. This is exactly the protocol of continuous variable quantum erasing as mentioned above. In most systems, however, it is not possible to squeeze the environmental modes which are normally inaccessible to the experimenter. These modes are either vacuum or thermal states which means that a full recovery of the quantum state after transmission in a noisy environment is not possible by means of environmental measurements and classical feedforward. However, as we will show in this paper it is possible to fully recover the quantum information carried by the quadratures of a coherent state even though the environmental modes are vacuum or thermal. The method developed in this paper we coin {\em environmental assisted quantum information correction}, and it basically corresponds to phase-insensitive quantum erasing in contrast to the "standard" quantum erasing approach which is phase-sensitive. A similar method was treated theoretically in ref.~\cite{Filip2},to eliminate a mode crosstalk in the context of signal multi-plexed quantum key distribution.

The quantum information considered in this paper is encoded as pure coherent states which undergo a noisy and lossy evolution in a simulated environment. We show that by placing a heterodyne detector in the environment, it is possible to retrieve sufficient information to {\em fully and deterministically} regain the continuous variable quantum information simultaneously carried by conjugate quadratures of the transmitted state. This holds true, however, only under certain conditions: First, one needs to know the losses in the channel, and second, one needs to have complete access to the environmental leakage modes using ideal heterodyne detection.
Complete access to the information carrying modes of the environment is normally not possible in a realistic setting.
However, we show that even if the access to the information modes is meniscule, it is still possible to fully remove the {\em excess noise} of the channel. The amount of excess noise is of high importance for establishing the security of a communication channel, so its removal based on a very unsharp environmental measurement is a promising technique to ensure security of noisy channels.

As an alternative to the deterministic approach we also present a probabilistic approach where the conveyed states are probabilistically selected based on the meansurement outcomes in the environment. The main advantage of this approach is that no prior information about the losses in the channel is required to execute the correction.  Such a probabilistic protocol will be also demonstrated in the paper.

Finally, in this paper we investigate theoretically the use of our protocol in a quantum key distribution system that is based on coherent state encoding and heterodyne measurements. We find that a security breaking channel (due to excess noise) can be turned into a security preserving channel by exploiting partial information of the environmental modes.   

\section{The system}

\begin{figure}[h]
 \centering
  \includegraphics[width=8cm]{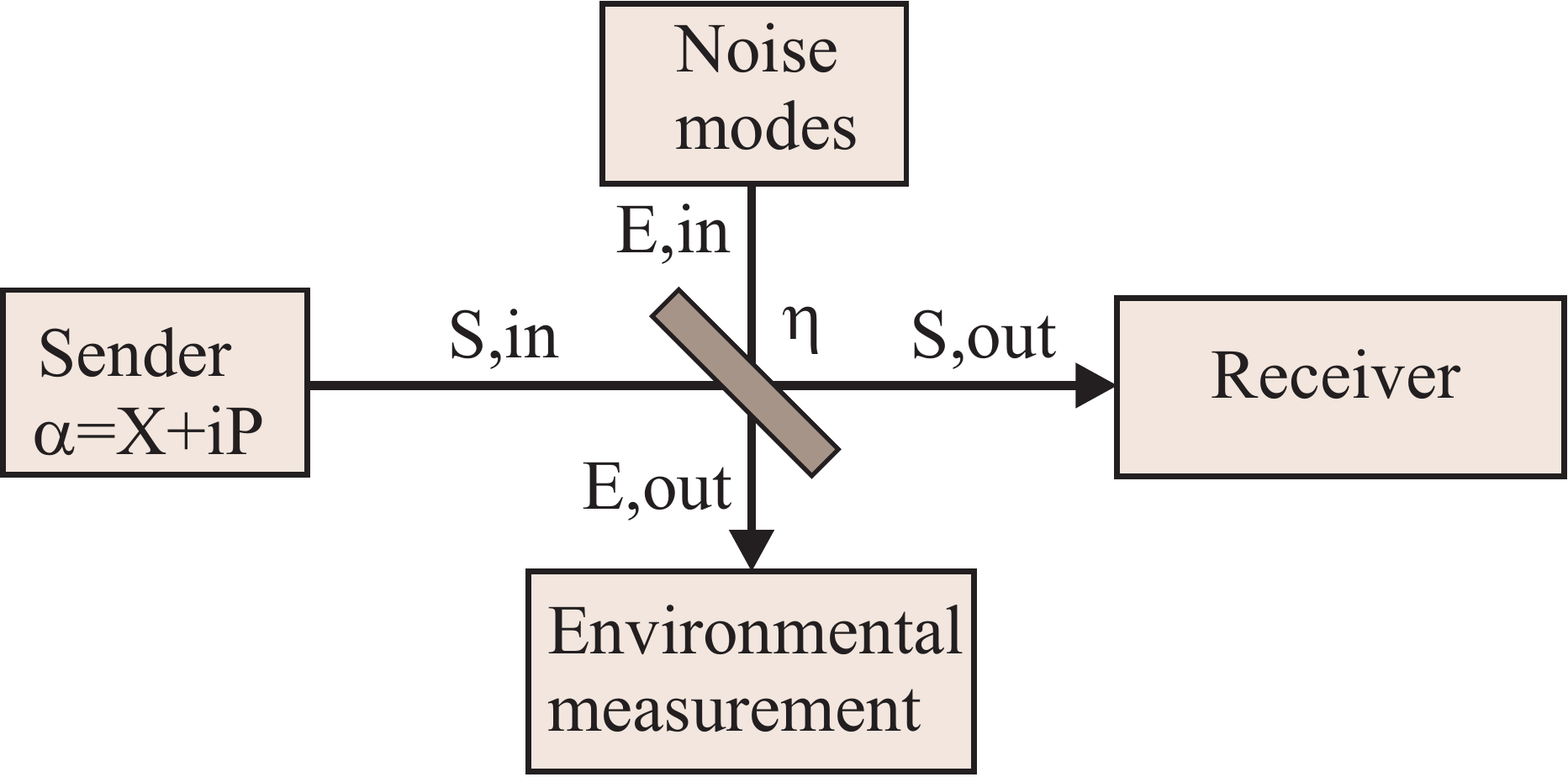}\caption{Schematic of the investigated protocol. Sender: Information is encoded in conjugate quadratures (or in a single quadrature) of a coherent state. Environment: The environment is simulated by a variable beam splitter with injected Gaussian noise, and the environmental measurement is carried out with a heterodyne detector (or homodyne detector). Receiver: The information is measured with a heterodyne (or homodyne) detector.}
 \label{fig:scheme}
\end{figure}

Let us consider the standard coherent state communication system depicted in Fig.~\ref{fig:scheme}. Information is encoded into conjugate quadratures of the coherent state of an electromagnetic field; the amplitude, $X$, and the phase quadrature, $P$, obeying the commutation relation $[X,P]=2i$. At the receiving station, information is efficiently measured using either homodyne or heterodyne detectors corresponding to a sharp measurement of a single quadrature or unsharp but simultaneous measurements of conjugate quadratures. Between the sender and receiver stations, the information carrying coherent state is conveyed in a noisy quantum channel. The decoherence process, which is assumed to be linear, is simulated by a beam splitter that couples the signal, denoted $S$, with the noisy environmental modes, denoted $E$. The quadratures of the noisy modes are assumed to have Gaussian statistics, thus the output modes of the channel is likewise Gaussian.
Let us note again that by tracing out the environmental modes, the excess noise cannot be removed solely using Gaussian operations. (Note that this only holds true for single copies. For multiple copies it is possible through interference to reconstruct the original coherent state if only excess noise has been added to the state~\cite{andersen05.pra}).

The unitary coupling between the signal and the environment is in the Heisenberg picture written as
\begin{eqnarray}
X_{S,out}=\sqrt{\eta}X_{S,in}+\sqrt{1-\eta}X_{E,in}\\
P_{S,out}=\sqrt{\eta}P_{S,in}+\sqrt{1-\eta}P_{E,in}\\
X_{E,out}=\sqrt{1-\eta}X_{S,in}-\sqrt{\eta}X_{E,in}\\
P_{E,out}=\sqrt{1-\eta}P_{S,in}-\sqrt{\eta}P_{E,in}
\end{eqnarray}
where $\eta$ is the coupling strength (or channel transmission). As a result of the noisy coupling to the environment the transmitted signal mode will be infected by added noise with a variance given by
\begin{equation}
V_{add}=\frac{1-\eta}{\eta}V_{E,in}
\label{add}
\end{equation}
where $V_{E,in}$ is the variance of enviromental input Gaussian modes.
If the variance of the excess noise, given by $\epsilon=(1-\eta)(V_{E,in}-1)/\eta$, is larger than 2 vacuum units the channel is not entanglement preserving and therefore will be insecure against any eavesdropping attacks. Moreover, to withstand the more powerful collective attacks the excess noise should be less than 0.8 vacuum units, thus putting more stringent conditions on the channel performance. In the next section we discuss how the added noise and the excess noise can be reduced by measuring the environmental modes.

\section{Environmental assisted channel correction}
\label{theory}

We now assume that we have access to the environmental modes through classical measurements. As mentioned in the introduction, it has been shown that for qubit and qutrit systems it is possible to fully recover the quantum state through measurements and classical feedforward~\cite{Buscemi1}. For our continuous variable system embedded in environmental vacuum or thermal modes, we will not be able to recover the coherent state through measurements, but it is possible to recover the quantum information encoded in the state as we will show in the following.

We consider two different environmental measurements; the homodyne and the heterodyne measurement. Since it is unrealistic in practice to have complete access to the leaky mode (described by $X_{E,out}$ and $P_{E,out}$), in our model we introduce additional loss to the mode before measurement. The quadratures being measured in the environment is therefore given by
\begin{eqnarray}
X_{E,mea}=&&\sqrt{\Gamma}\left(\sqrt{\gamma}X_{E,out}+\sqrt{1-\gamma}X_{vac1}\right)\\
&+&\sqrt{1-\Gamma}X_{vac2}\\
P_{E,mea}=&&\sqrt{1-\Gamma}\left(\sqrt{\gamma}P_{E,out}+\sqrt{1-\gamma}P_{vac1}\right)\\
&-&\sqrt{\Gamma}P_{vac2}
\end{eqnarray}
where $\gamma$ represents the fraction of the leaky mode being measured and $X_{vac1,2}$ and $P_{vac1,2}$ are vacuum noise operators. $\Gamma$ is a discrete function associated with the two different detectors:
\begin{equation}
\Gamma=\left\{\begin{array}{cc}
1,\mbox{Homodyne detection}\\
1/2, \mbox{Heterodyne detection}\end{array}\right.
\end{equation}
Assuming the electronic noise of the detectors to be negliable, the measurement will result in some classical data: $X_{E,mea}\rightarrow x_{E,mea}$ and $P_{E,mea}\rightarrow p_{E,mea}$. This information is subsequently used to restore the degraded signal by means of classical feedforward: The measurement outcomes are scaled by an electronic amplifier and imposes a phase space displacement onto the signal. If the information is encoded in a single quadrature, say the amplitude quadrature, the environmental measurement should be a homodyne measurement that performs a sharp amplitude quadrature measurement. On the other hand, if information is encoded in conjugate quadratures, the measurement in the environment is a heterodyne detector where both quadratures are detected simultaneously, but unsharply.

\subsection{Single quadrature encoding}
We first consider the case where information is encoded into one quadrature, the amplitude quadrature. The single quadrature information leaks into the environment and is partly measured with a homodyne detector. This results in the outcome $x_{E,mea}$ which is then used to displace the state transmitted through the channel: $X_{S,out}\rightarrow X_{S,out}+g_x x_{E,mea}$ where $g_x$ is the the electronic gain of the classical feedforward loop. By choosing $g_x=\sqrt{(1-\eta)/(\gamma\eta)}$ the noisy environmental mode $E$ can be completely decoupled from the signal and the signal itself will be amplified, thus resulting in the following output
\begin{equation}
X_{S,out}=\sqrt{G}X_{S,in}+\sqrt{\alpha_{Hom}}\sqrt{G-1}X_{vac},
\label{xsout}
\end{equation}
where $G=1/\eta$ is the optical gain and
$\alpha_{Hom}=(1-\gamma)/\gamma$ corresponds to the normalized detection efficiency
with which mode $E$ is measured. It is worth noting that the noisy environmental mode is perfectly removed independent of the efficiency of the measurement. In other words, if only a tiny part of the environment is accessible to our homodyne detector, it is still possible to completely remove the noisy environmental modes assuming zero electronic noise. The noisy modes are however substituted with vacuum modes which therefore results in the added noise
\begin{equation}
V_{add}^{Hom}=\frac{(1-\eta)(1-\gamma)}{\gamma}.
\label{addHom}
\end{equation}
This tends to zero when the environmental measurement is perfect.
By comparing (\ref{addHom}) with (\ref{add}) we see that the measurement induced correction is only effective if $\gamma>\eta/(V_{E,in}+\eta)$. To remove the excess noise of the environmental modes, the variance of this noise, , the environmental detector efficiency $\gamma$  as well as the coupling strenght, $\eta$, must be a-priori known (since the feedforward gain $g_x$ is a function of these parameters). If the coupling strength, the environmental detection efficiency  and the noise variance are stationary parameters, they can be estimated using a serie of probe pulses before the actual signal is sent. However, if they are non-stationary, estimation is only possible with measurements that are much faster than the environmental changes.

Note that the feedforward system (without the noisy modes) is identical in operation to the one proposed and implemented by Lam et al.~\cite{lam98} and Buchler et al.~\cite{buchler98} to enable non-unitary noise-less amplification of the amplitude or phase quadrature, respectively. This is easily seen from equation (\ref{xsout}) since for ideal detection, $\gamma=1$ and $X_{S,out}=\sqrt{G}X_{S,in}$ which exactly represents the input-output relation for a noiseless quadrature amplifier. It should also be noted that the noiseless amplifier can be made unitary by injecting squeezed light into the beam splitter (instead of a noisy mode as above or a vacuum mode as in refs.~\cite{lam98,buchler98}) as proposed in ref.~\cite{filip05} and implemented in ref.~\cite{ichi07}. 


\subsection{Conjugate quadrature encoding}
In most CV quantum communication systems, information is encoded in conjugate quadratures; the alphabet of input states is in many cases a symmetric Gaussian distribution of coherent states. For such communication scenarios, the environmental noise will be detrimental to conjugate quadratures and the question is thus whether this noise in conjugate quadratures can be simultaneously removed. Now, instead of using a homodyne detector, we employ a heterodyne detector ($\Gamma =1/2$) which yields two outcomes; $x_{E,mea}$ and $p_{E,mea}$ which are used to displace the signal state to
\begin{eqnarray}
X_{S,out}&=&\sqrt{G}X_{S,in}+\sqrt{\alpha_{Het}}\sqrt{G-1}X_{vac}\\
P_{S,out}&=&\sqrt{G}P_{S,in}+\sqrt{\alpha_{Het}}\sqrt{G-1}P_{vac}
\label{amplifier}
\end{eqnarray}
where the electronic gains are set to $g_X=g_P=\sqrt{2(1-\eta)/(\gamma\eta)}$ and the optical gain is $G=1/\eta$ and $\alpha_{Het}=(2-\gamma)/\gamma$. Just as in the single quadrature case the noise of the environmental modes is perfectly decoupled from the signal as a result of the measurement induced displacement. However, in contrast to the single quadrature case, the signal will be phase in-sensitively amplified which means that noise will be added to the signal no matter how perfect the environmental measurement is carried out. It is clearly seen from the input-output relations in eqn. (\ref{amplifier}) that for perfect detection in the environment (corresponding to $\gamma=1$), the scheme resembles an ideal phase insensitive amplifier similar to the one proposed and implemented in ref.~\cite{josse06}. For an arbitrary efficiency of the environmental measurement, the variance of the added noise to the quantum state is 
\begin{equation}
V_{S,add}^{Het}=\frac{(1-\eta)(2-\gamma)}{\gamma}
\label{addHet}
\end{equation}
To reduce the added noise of the state by feedforward correction, the detection efficiency $\gamma$ should satisfy the condition; 
\begin{equation}
\gamma>\frac{1}{1+\frac{V_{E,in}}{\eta}}.
\end{equation}  
It is interesting to consider two limiting cases corresponding to an almost additive noise channel (associated with large $\eta$ and $V_{E,in}$) and the highly lossy channel with little excess noise(associated with small $\eta$ and $V_{E,in}$). For $\eta\approx 1$ and $V_{E,in}$ being very large, the added noise before correction is very large, but the corrective action reduces this noise very drastically even when the efficicency of the environmental measurement is small. This is simply due to the large classical correlations between the environmental modes and the signal modes. On the other hand, for small $\eta$ and $V_{E,in}$, the condition on detection efficiency is rather tight. 

Although the added noise is reduced as a result of the feedforward action, it can never go to zero. This means that the original quantum state (described in the infinitely dimensional Hilbert space) cannot be perfectly reconstructed. 
However, it turns out that it is possible to fully get back the information encoded in the state using the measurement induced operation. In the case where information is decoded by applying a heterodyne detector at the receiving end of the communication link, the environmental measurement aids a full recovery of the information. In other words, by combining the measurement results of the conjugate quadratures in the environment and the receiving station, perfect information retrieval is obtained, thus resembling the ideal and loss-free channel.
The added noise for both conjugate quadratures in the heterodyne measurement of the signal at the receiving station is given by (without feedforward)
\begin{equation}
V_{S,add}^{wo ff}=\frac{(1-\eta)V_{E,in}+1}{\eta}
\label{vhet'}
\end{equation}
whereas with the feedforward, it changes to
\begin{equation}
V_{S,add}^{w ff}=\eta+\frac{(1-\eta)(2-\gamma)}{\gamma}
\label{vhet''}
\end{equation}
which is smaller than (\ref{vhet'}), if
\begin{eqnarray}
 \gamma>\frac{2\eta}{1 + 2 \eta + V_{E,in}}
 \end{eqnarray}
Remarkably, we see that for ideal heterodyning ($\gamma=1$),
the added noise after feedforward correction is only a {\em single} vacuum unit. This is exactly the same amount of added noise for heterodyne detection that would be expected for an ideal lossless and noiseless channel; by use of a destructive measurement and {\it classical} feedforward, the {\it quantum} information has been perfectly recovered, thus allowing ideal coherent state communication using conjugate quadratures.
Thus, even though the quantum state cannot be reconstructed using classical feedforward,
the information content carried by the quantum state which was buried under noise can be recovered perfectly.

In the above scheme the electronic feedforward gain was chosen to completely erase the noisy environmental modes. However, this choice of the gain does not correspond to a minimization of the added noise. The minimal value of the added noise of the quantum state is
\begin{equation}
V_{S,add}^{opt}=\frac{(1-\eta)(2-\gamma)V_{E,in}}{\eta (2-\gamma)+\gamma V_{E,in}}
\label{Vopt}
\end{equation}
which is obtained for 
\begin{equation}
g_X=g_P =\frac{\sqrt{2\gamma(1-\eta)}V_{E,in}}{\sqrt{\eta}(2+\gamma(V_{E,in}-1))} 
\end{equation}
corresponding to an optical gain of
\begin{equation}
G=\frac{1}{\eta}\left(\frac{(2-\gamma)\eta + \gamma
V_{E,in}}{2-\gamma+\gamma V_{E,in}}\right)^2
\label{gain}
\end{equation}
Note that the minimized added noise after feedforward (eq. (\ref{Vopt})) is always smaller than the added noise that is presence without feedforward (eq. (\ref{add})). However, in contrast to the feedforward strategy resulting in the added noise in eq. (\ref{addHet}), here the added noise depends on the excess noise.

\subsection{Probabilistic approach}
The drawback of all the above mentioned feedforward correction strategies is the fact that one needs a-priori knowledge about the coupling strenght, $\eta$, to the environment and in the last case also knowledge about the excess noise of the channel modes in order to deterministically recover the information content of the signal. If however one uses a probabilistic strategy, the knowledge about the environment can be relaxed. Such a method is implemented by replacing the linear feedforward loop with a triggering loop detector: the measurement outcome of the environmental detector determines whether the signal should be kept or discarded. If the outcome lies within a pre-specified interval (defined as $[-X_{th},X_{th}]$), the signal should be kept, otherwise it should be discarded. The feedforward gain is thus a binary function and therefore independent of the environmental parameters. The drawback of this method, however, is its probabilistic nature: As the postselection interval is getting narrower (corresponding to a more efficient information recovery) the success rate decreases.

To calculate the result of the probabilistic method, we employ the covariance matrix formalism \cite{CM}. The first and second moments of the environmental mode and signal mode after the beam splitter coupling are in this formalism collected in the vectors $D_S$ and $D_{E}$ and the 4x4 covariance matrix $CV_{ES}=\left((A,C),(C^T,B)\right)$. The heterodyne measurement of the environmental mode and subsequent heralding of the signal mode corresponding to the measurement outcomes $x_{E,out}=0$ and $p_{E,out}=0$ is described by the following transformation 
\begin{eqnarray}
CV_{ES}\rightarrow CV_{S,add}^{prob}=B-C\frac{1}{A+M}C^T\\
D_S\rightarrow D_{S}^{prob}=D_S+C\frac{1}{A+M}(D_M-D_{E})
\end{eqnarray}
where $M=((1,0),(0,1))$ and $D_M=(0,0)$ for ideal heterodyne detection. 
In this limit, the added noise after heterodyne detection at the receiver is
\begin{equation}
V_{S,add}^{prob}=\frac{1-\eta}{\eta}\frac{(1-\gamma)\gamma\eta
(V_{E,in}-1)^2+V_{E,in}}{(1+\gamma (V_{E,in}-1))^2}
\end{equation}
and the channel gain is 
\begin{equation}
G^{prob}=\frac{\eta(1-\gamma+\gamma V_{E,in})^2}{(1-\gamma+\gamma(\eta
V_{E,in}+(1-\eta)))^2}.
\end{equation}
The added noise is now a function of the
variance $V_{E,in}$ for any $\gamma$ in contrast to the 
deterministic methods. Using the probabilistic approach it is therefore impossible to perfectly decouple the noisy mode $E$ from the signal mode $S$. However, for any $\gamma>0$ and $V_{E,in}>1$, the resulting added noise is always smaller than would be obtained without any correction (given by eq. (\ref{add})). 


In the above derivation, it was assumed that $x_{th}=p_{th}=0$ which will yield a vanishing small success probability. To estimate the added noise for which a range of measurement outcomes yield success, numerical approach based on Wigner function analysis have been employed. The results of these calculations are presented in following sections on the experimental implementation.  

\section{Experimental setup}

We now proceed with an experimental demonstration of environmental assisted quantum information correction in the case where information is encoded in conjugate quadratures of a coherent state. The experimental setup is shown in figure 2 and it consists of three main parts; a sender station where quantum information is encoded, a noisy quantum channel, an environmental measurement and finally a receiver station in which the conveyed information is measured.

Preparation of the coherent states is accomplished by employing an amplitude and a phase modulator placed in the beam path and driven by function generators modulating at 14.3 MHz. The modulators create pure sideband frequency modes with easily controllable excitations (or displacements in phase space). The input quantum state is thus defined as radio-frequency sidebands to the optical carrier. 

\begin{figure}[!h]
 \centering
  \includegraphics[width=8cm]{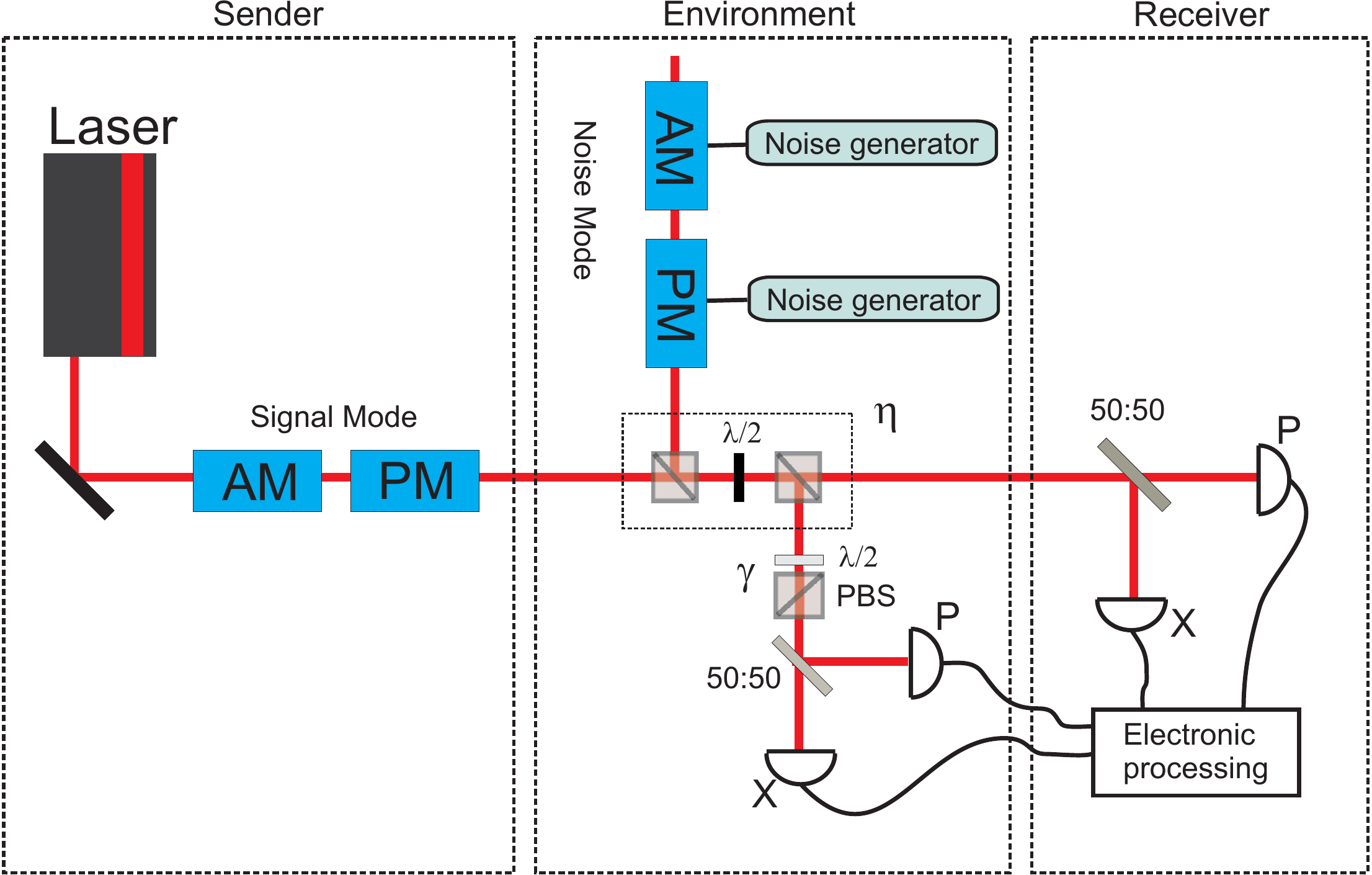}\caption{Schematic of the experimental setup. Information is encoded by using an Amplitude Modulator (AM) and a Phase Modulator (PM) operating at 14.3MHz. The beam splitter coupling is controlled by combining two PBS with a half-wave plate, and the efficiency of the environmantal measurement is likewise controlled by a PBS and a half-wave plate. The environmental noise modes are generated by two modulators fed by indpendent Gaussian noise. Measurements are carried out by simultaneously detecting conjugate quadratures as described in the main text.} 
 \label{fig:setup}
\end{figure}

\begin{figure*}[!t]
 \centering
  \includegraphics[width=18cm]{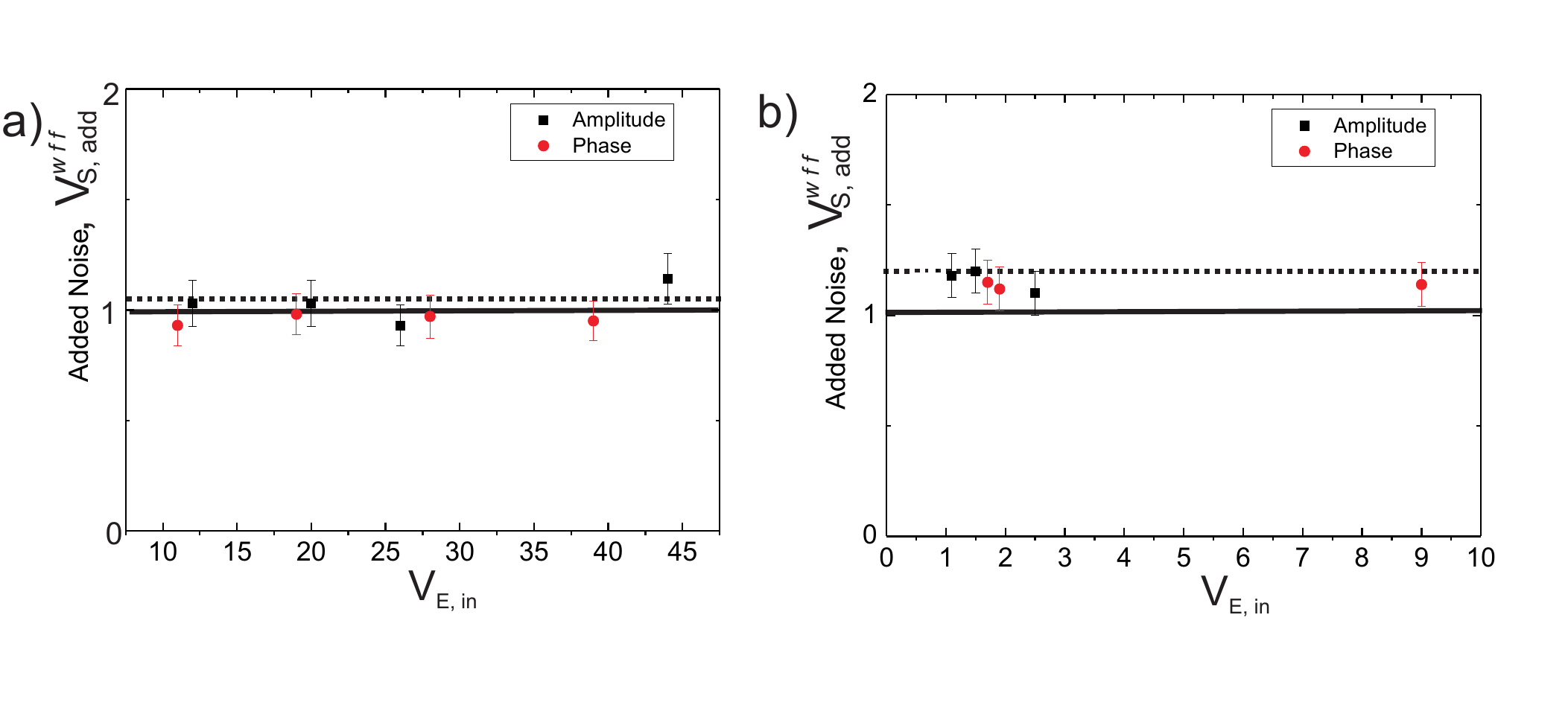}\caption{The results for the deterministic approach. The added noise, $V_{S,add}^{w\;ff}$, is plotted against the amount of excess noise from the environment $V_{E,in}$. a) The weak  coupling regime with $\eta=0.9$ b) The strong coupling regime with $\eta=0.1$  The solid line is associated with the ideal noise cancellation, and the dashed line takes into account the measurement imperfections in the environment. The non-unity measurement efficiency of the environment plays a bigger role for the strong coupling regime, due to the fact that more information will in this case leak into the environment. Without employing the corrective action, the added noise ($V_{S,add}^{wo\;ff}$) would have been between 2.2 and 6.1 for $10<V_{E,in}<45$ in the weak coupling regime and between 19.9 and 91 for $1.1<V_{E,in}<9$ in the strong coupling regime.}
 \label{fig:feedforward}
\end{figure*}

The coherent state couples to the noisy environmental modes through a variable beam splitter composed of a half wave plate sandwiched between two polarizing beam splitters (PBSs). The signal mode and the noisy mode enters the first PBS in orthogonal polarizations and interfere at the second PBS with a ratio determined by the orientation of the phase plate. We can therefore easily tune the coupling strength (channel losses) by a simple phase plate rotation. The noise modes are prepared also at the rf sidebands by traversing an auxiliary beam through an amplitude and a phase modulator driven by two independent electronic noise generators. The noise is white over a frequency range which is much broader than the measurement bandwidth. The amount of excess noise (or the variance, $V_{E,in}$) is easily adjusted through the voltage output of the electronic noise sources.

Measurements in the environment was carried out by a high efficiency heterodyne detector. To implement the measurement we make use of an auxiliary beam with the same brightness as the signal. The two beams interfere on a 50/50 beam splitter with a relative phase of $\pi/2$ and the two resulting outputs are measured with two low noise detectors. Joint measurements of $X$ and $P$ are thus obtained by electronically constructing the sum and the difference currents. The total quantum efficiency of the detector (including photodiode efficiency (95\%) and interference contrast (97\%)) is about 90\%. However, in order to simulate a less efficient measurement in the environment we place a controllable attenuator (half-wave plate and a PBS) in front of the detector, thus controlling the parameter $\gamma$. The heterodyne measurement at the receiving station is carried out similarly to the one of the environment, but its efficiency is lower (about 80\%) due to a worse interference contrast (94\%) and diodes with less efficiency (90\%).

The environmental information obtained by the heterodyne detector is then used to correct the transmitted state. The correction can be carried out optically using two electro-optical modulators controlled by the environmental measurements or electronically by linearly displacing (for the deterministic approach) or "chopping" (for the probabilistic approach) the photocurrents of the receiver detectors depending on the measurement outcomes. We chose the latter approach due to its simplicity.

In the following we use the described setup to perform environmental assisted channel correction using either the deterministic or the probabilistic approach.

\subsection{Deterministic approach}
As mentioned above, when applying the deterministics approach for correcting errors, information about the coupling strength must be a priori known in order to implement the right feedforward gain. We investigate two transmission scenarios corresponding to $\eta=0.1$ and $\eta=0.9$. The optical gains after feedforward for these channels  should be $G=10$ and $G=1.11$, respectively.

To characterize the protocol, we measure the first and second order moments of the input and output state using a spectrum analyzer. Note that due to the Gaussian statistics of $X$ and $P$, the first two moments fully characterise the states. The spectrum analyzer is used in
zero span mode at the frequency of 14.3~MHz, a resolution bandwidth of 10kHz and a video bandwidth of 30 Hz. We set the electronic gain of the feedforward loop such that the calculated optical gain is correctly obtained which is confirmed by the measurements of the first moments using the spectrum analyzer. Now when the gain is set, we measure the noise power using the spectrum analyzer and use these data to calculate the signal-to-noise ratios and subsequently the added noise variances. 

The added noise variance associated with heterodyne detection is found for different amounts of excess noise in the environment, and the results after the corrective action are summarized in Fig.~\ref{fig:feedforward}. In Fig.~\ref{fig:feedforward}a and b the coupling strenghts are set to $\eta=0.9$ and $\eta=0.1$, respectively. It is clearly seen that the feedforward correction loop removes the excess noise almost completely, thus the added noise nearly attains the value of a single vacuum unit as expected for ideal heterodyne detection (and marked by the solid horizontal line). The expected values taking into account the inefficiency of the environmental heterodyne detector are found to be $V_{S,add}^{w\;ff}= 1.16$ and $V_{S,add}^{w\;ff}= 1.02$  for $\eta=0.1$ and  $\eta=0.9$, respectively. These theoretical values are marked by the dashed horizontal lines, and fits reasonably well with the measured data within their error bars.

\begin{figure}[h]
 \centering
  \includegraphics[width=9cm]{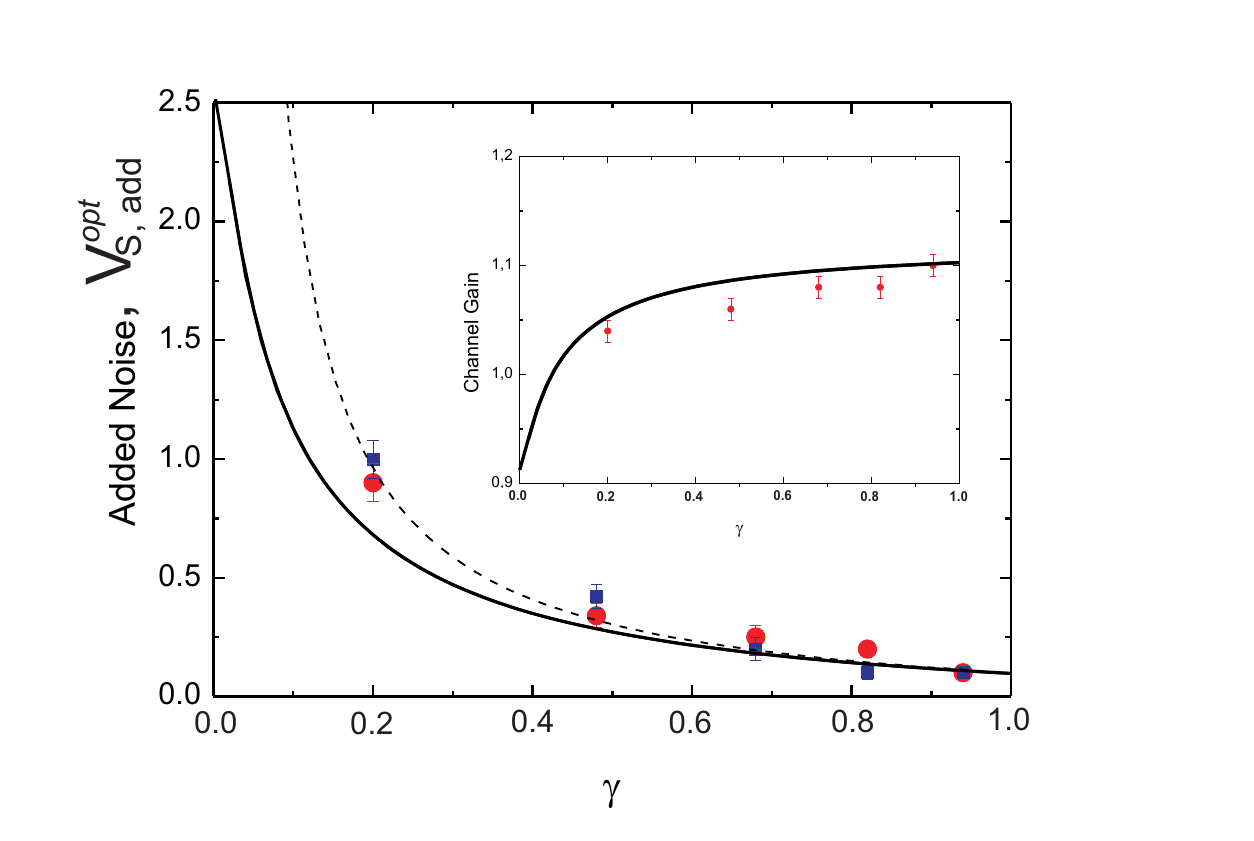}\caption{The added noise, $V_{S,add}^{opt}$, is plotted against the efficiency of the environmental measurement, $\gamma$. The blue squares and red dots correspond to the amplitude and phase quadrature measurements respectively. The solid and the dashed lines correspond to the theoretical prediction for the optimised approach with respect to noise minimization (eq. (\ref{Vopt}) with $V_{E,in}=25$ and $\eta=0.9$) and the approach for which the excess noise is removed (eq. (\ref{addHet})). The added noise when measured directly without any correction is $V_{add}=2.77$. The inset shows the channel gain as a function of $\gamma$, and the solid curve is associated with theory (eq.~(\ref{gain})).}
 \label{fig_efficiency}
\end{figure}

As mentioned in the theory section, the feedforward gain associated with the complete removal of the excess noise is not neccessarily coinciding with a minimization of the added noise. For the minimization of the added noise, the electronic feedforward gain is optimised with respect to the coupling strength, the measurement efficiency of the environmental measurement and the excess noise in the environment. In the experiment we vary the environmental measurement efficiency while fixing the coupling strength and the variance of the excess noise. 
The measurements of the added noise of the state (using homodyne detection) is shown in Fig.~\ref{fig_efficiency} and the measurements of the optical gain is shown in the inset of Fig.~\ref{fig_efficiency}. The solid curves are the theoretical predictions from the theory presented in section \ref{theory}. For comparison we insert the expected added noise(dashed curve) for the case where the electronic feedforward gain is set to remove the excess noise (but not minimize the added noise) corresponding to eqn. (\ref{addHet}). The discrepancy between the experimentally obtained results and the theoretical solid curve for low transmissions (small $\gamma$) is due to the increasing relative electronic noise of the detectors for lower optical powers. This increasing electronic noise was caused by the actual heterodyne measurement which required the auxiliary beam to be attenuated by the same amount as the signal beam in order to access conjugate quadratures.


\subsection{Probabilistic approach}
We now turn our attention to the experimental investigation of the probabilistic scheme. As mentioned above, the probabilistic feedforward correcting operation is independent of the channel parameters; the measurement outcomes solely determine whether the transmitted state should be discarded or kept.

\begin{figure*}[!t]
 \centering
  \includegraphics[width=18cm]{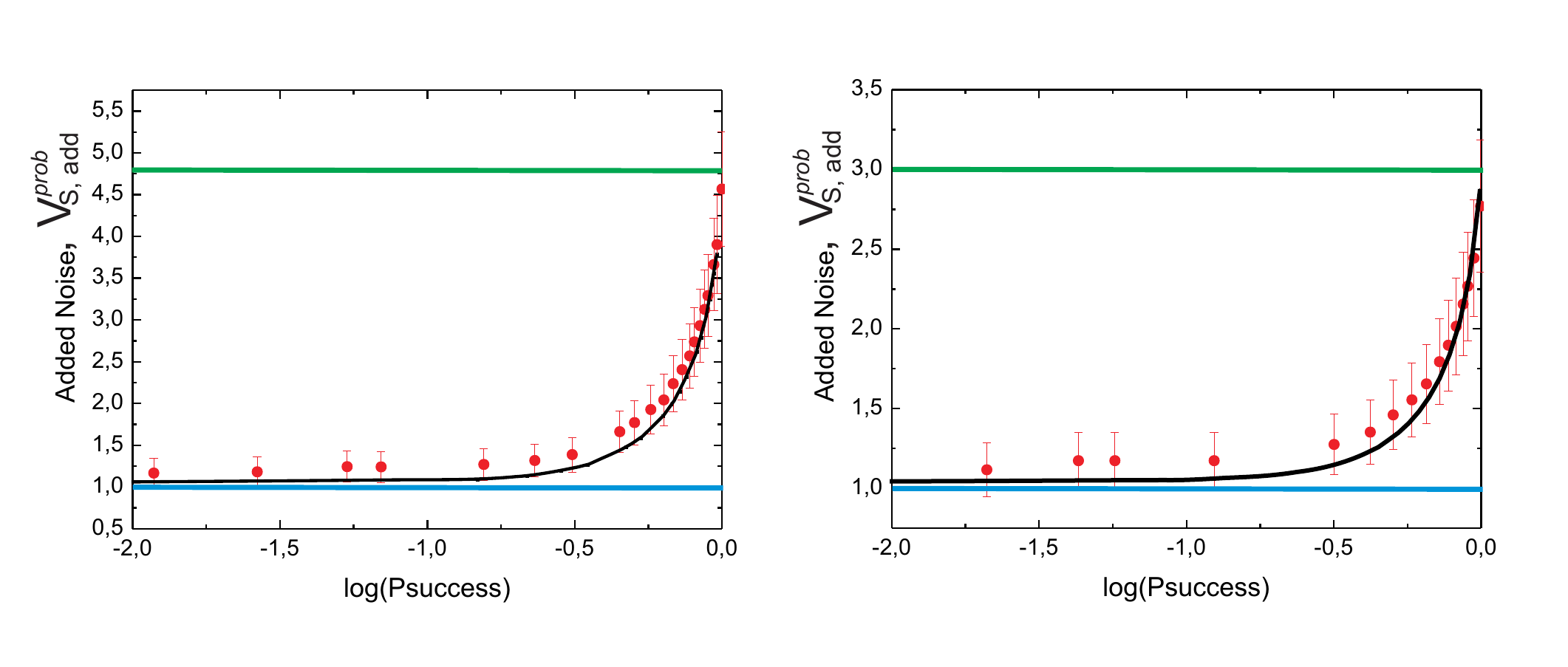}\caption{The added noise of the receiving heterodyne detector as a function of the success probability for the amplitude quadrature (left figure) and the phase quadrature (right figure). The solid curve corresponds to numerical calculations using Wigner function analysis, and the upper straight lines are the added noises without feedforward corrections. These variances are $V_{S,add}^{wo\;ff}=4.55$ and $V_{S,add}^{wo\;ff}=3$ for the left and right figures, respectively. The lower lines correspond to perfect noise cancellation.}
 \label{fig:probamp}
\end{figure*}

To implement such a heralding process, the measurements are carried out in time-domain rather than in frequency domain in order to access the actual measured quadrature amplitudes. The radio frequency outputs from the detectors are downmixed using an electronic mixer with a strong electronic local oscillator centered at 14.3~MHz. The downmixed signal is then amplified, low pass filtered with a cutoff frequency of 150 kHz and finally recorded on a computer (10 million samples recorded at 5Ms/s). We then have two sets of data; a set from the environmental detector and a set from the receiver detector. The data from the receiver is subsequently postselected based on the data from the environment; if the $X$ and $P$ values from the environment are smaller than a certain threshold value, the corresponding data pair in the receiver data set is kept, otherwise it is discarded. From the remaining receiver data we calculate the added noise and the success probability (ratio of the data kept after postselection to all data) for the amplitude and the phase quadrature. This is then done for various threshold values and the results are illustrated in Fig.~\ref{fig:probamp} as a function of success probability for two different excess noise variances. For large threshold values (corresponding to large success probability), nearly all data are kept and the added noise is large. However, as the threshold value decreases, the pure coherent states are selected from the mixture and the added noise approaches the optimal of one shot noise unit.

\section{Application to Quantum Key Distribution}

Modulated coherent states and homodyne (heterodyne) detectors can be used to implement quantum key distribution (QKD)~\cite{Grosshans2002.prl,Silberhorn2002.prl,Weedbrock}. In many QKD protocols, information is carried by states from a coherent state alphabet and the receving station measures conjugate quadratures; either serially in a random sequence using homodyne detection or simultaneously using heterodyne detection. After all measurements have been completed, the receiver and sender share a common set of correlated data from which a potential secret key can be drawn using the techniques of either direct or reverse reconciliation. To establish secret keys in a lossy channel, reverse reconciliation is used since it is more robust against loss. Such a protocol based on Gaussian modulated coherent states and reverse reconciliation has been implemented~\cite{Grosshans2003.nat} and the impact of noise on the security of the channel has been addressed by considering the collective attack scenario~\cite{c1}. Furthermore, it was proved that 
known preparation~\cite{filip2008.pra} and detection noise~\cite{Lodewyck2007.pra} do not break the security of the channel. On the other hand, excess noise in the channel can significantly reduce the secure key rate 
and above a certain noise-level the channel is no longer secure and thus security breaking. 
If the channel is still entanglement preserving, continuous-variable quantum repeaters can be 
used to re-establish secure communication. But if the channel is entanglement breaking, quantum repeaters cannot be used and the channel has to be repaired or replaced to allow for secure communication. Alternatively, if one has access to some of the environmental modes it is actually possible to regain security using the method of environmental assited channel correction. 
  
To illustrate this, we consider a specific QKD scheme based on Gaussian modulated coherent states and random homodyne detection. We assume that the transmission of the channel is 90\% and the added noise is $2.77$. The excess noise of this channel is $\epsilon=\Delta V-(1-\eta)/\eta=2.67>2$ and, therefore, it is clearly 
entanglement and security
breaking~\cite{c1}. Using data from the experiment where the gain was optimised to minimize the added noise (see Fig.~4), we calculate expected secure key rate after the correction. 
The secure key rates for homodyne detection and direct reconcilliation method are 
presented in Table~1. Direct reconciliation is preferable for the amplifying channel [30] as is the case after implementation of the corrective action. 
The secret key rates for conjugate quadratures $K_X$ and $K_P$ have been obtained from the measured data depicted in Fig.~4 and calculated for a Gaussian modulation variance of $\sigma=40$. In the parenthesis, 
we state the maximal secure key rate is for $\sigma\rightarrow\infty$. These numbers can be compared to the optimal secure key rate $K_{opt}$ using feedforward correction (using Eqs.~(19,20) with $\eta=0.9$ and $V_{E,in}=25$). For $\gamma=1$, the maximal secure key rate is 
$K_{max}=1.443(1.617)$ if we minimized the added noise. 
Note that by using reverse reconcilliation in replacement of direct reconciliation, the secure key rate is much lower. As an example we consider the case where the environmental measurement has an efficiency $\gamma=0.48$, which yield the rates $K_X=0.012 (0.047)$ and $K_P=-0.177(-0.150)$ based on the data in Fig.~4 and reverse reconciliation. For ideal correction the rate is $K_{opt}=0.142(0.184)$. 


\begin{table}
	\centering
	\small
		\begin{tabular}{c c c c c c c}
		\hline
		$\gamma$ & $V^{opt,X}_{S,add} $ & $V^{opt,P}_{S,add}.$ & Ch. Gain &  $K_X$ & $K_P$  & $K_{max}$\\
		\hline
0.92 &     0.1  &    0.1&       1.1 &  1.38(1.53) & 1.38(1.53)  & 1.27(1.39)\\    
0.82  &    0.27  &   0.17&      1.08 &  0.65(0.69) & 0.93(1.00)  & 1.10(1.19)\\
0.68   &   0.27   &  0.32 &     1.08  & 0.65(0.69)  &0.54(0.58)   &0.90(0.97)\\
0.48    &  0.34    & 0.42  &    1.06   &0.50(0.53)  &0.38(0.40)   &0.62(0.65)\\
0.2      & 1.04     &0.94   &   1.04  &-0.17(-0.17) &-0.11(-0.11) &0.09(0.10)\\
\hline
	\end{tabular}
\caption{Theoretical secure key rates of quantum key distribution protocol with
coherent states and homodyne detection after the environmental assisted
quantum information correction.}
\end{table}


\section{Conclusion}
In conclusion, we have investigated a lossy and noisy quantum channel for which the environmental modes could be accessed by detectors. We found that for protocols involving coherent states and heterodyne detection, it is possible to fully recover the quantum information of the processed coherent state if a heterodyne detector efficiently measures the leaked environmental modes. The proposal has been experimentally implemented and we have successfully demonstrated the recovery of quantum information both deterministically and probabilistically.
Furthermore, we have investigated the use of environmental assisted quantum information correction in quantum key distribution. It was theoretically found that the method can be used to transform a channel from being security breaking to security preserving.  

This work is supported by the EU project COMPAS, the danish research council (FTP), the Lundbeck Foundation and the Deutsche Forschungsgemeinschaft. R.F. also acknowledge funding from projects No. MSM 6198959213 and No. LC06007 of the
Czech Ministry of Education, grant 202/07/J040 of
GA CR and the Alexander von Humbolt Foundation.

\end{document}